\documentclass[12pt]{article}
\usepackage{amsmath,amssymb,mathtools,cite,graphicx,fancyhdr,bbold,color}
\providecommand{\norm}[1]{\lVert#1\rVert}
\providecommand{\inner}[2]{\langle #1 | #2 \rangle}
\providecommand{\ket}[1]{\lvert #1 \rangle}
\providecommand{\bra}[1]{\langle #1 \lvert}
\providecommand{\conm}[2]{\left[ #1 , #2 \right]}
\providecommand{\esp}[1]{\langle #1 \rangle}
\newcommand{\MeijerG}[7]{G \begin{smallmatrix} #1\! & #2 \\ #3\! & #4 \end{smallmatrix}\!\! \left( \begin{smallmatrix} #5 \\ #6 \end{smallmatrix} \middle\vert #7 \right) }

\begin{document}
\lhead{Painlev\'e IV Coherent States}
\rhead{Bermudez, Contreras-Astorga, Fern\'andez C.}

\title{Painlev\'e IV Coherent States}
\author{David Bermudez\footnote{{\it email:} david.bermudez@weizmann.ac.il},\\
{\small Department of Physics of Complex Systems}\\ {\small Weizmann Institute of Science, Rehovot 76100, Israel}\\
Alonso Contreras-Astorga\footnote{{\it email:} aloncont@iun.edu } \\
{\small Department of Mathematics and Actuarial Science}\\ {\small Indiana University Northwest, 3400 Broadway, Gary IN 46408, USA}\\
David J. Fern\'andez C.\footnote{{\it email:} david@fis.cinvestav.mx}\\
{\small Departamento de F\'{\i}sica}\\ {\small Cinvestav, A.P. 14-740, 07000 M\'exico D.F., Mexico}
}

\date{}

\maketitle

\begin{abstract}
A simple way to find solutions of the Painlev\'e IV equation is by identifying Hamiltonian systems with third-order differential ladder operators. Some of these systems can be obtained by applying supersymmetric quantum mechanics (SUSY QM) to the harmonic oscillator. In this work, we will construct families of coherent states for such subset of SUSY partner Hamiltonians which are connected with the Painlev\'e IV equation. First, these coherent states are built up as eigenstates of the annihilation operator, then as displaced versions of the extremal states, both involving the third-order ladder operators, and finally as extremal states which are also displaced but now using the so called \emph{linearized} ladder operators. To each SUSY partner Hamiltonian corresponds two families of coherent states: one inside the infinite subspace associated with the isospectral part of the spectrum and another one in the finite subspace generated by the states created through the SUSY technique. 
\\

\end{abstract}

\section{Introduction}
In the dawn of quantum mechanics, Erwin Schr\"odinger \cite{Sch26} was interested in establishing a connection between the new science and classical mechanics. With this interest in mind, he found quantum states with the right classical behavior in phase space \cite{AAG00}. Since these states can be used to examine the behavior of several systems at the border between quantum and semi-classical regimes \cite{Kla63a,Kla63b,Per86,GK99,Que01,BHK12}, its study has taken an important place in quantum physics during the last fifty years. These are the \emph{coherent states} (CS), a term which was coined by Glauber when studying electromagnetic correlation functions \cite{Gla63a,Gla63b}. 

For the harmonic oscillator, the standard CS are expressed as 
\begin{equation}
\ket{z}=e^{-|z|^2/2} \sum_{n=0}^{\infty} \frac{z^n}{\sqrt{n!}} \ket{n}, \label{standard CS}
\end{equation}
where $\ket{n}$ are the normalized eigenstates of the harmonic oscillator Hamiltonian with eigenvalues $E_n=n+1/2$ (in dimensionless units) and $z \in \mathbb{C}$. Let us note that several properties of the standard CS are used as definitions to construct the corresponding states for other quantum systems, namely:

\begin{itemize}
\item The CS are eigenstates of the annihilation operator,
\begin{equation}
a^- \ket{z} = z \ket{z}, \qquad  z \in \mathbb{C}. 
\end{equation}

\item They arise from applying the displacement operator $D(z)$ onto the ground state $\ket{0}$,
\begin{equation}
\ket{z}=D(z) \ket{0}, \qquad D(z)= \exp \left(z a^+ - z^*a^- \right).
\end{equation}

\item Those states satisfy the minimum Heisenberg uncertainty relation for the position and momentum operators,
\begin{equation}
\left( \Delta X \right)_z  \left( \Delta P \right)_z = \frac{1}{2}.
\end{equation}

\item They allow to decompose the identity operator in the way
\begin{equation}
\mathbb{1}=\frac{1}{\pi} \int_\mathbb{C}\ket{z}\bra{z} \text{d}z.
\end{equation}
\end{itemize}
 
In this paper we will find sets of CS for what we will call {\it Painlev\'e IV Hamiltonian systems}, which are special families of $k$th order SUSY partners of the harmonic oscillator having associated always third-order differential ladder operators $l_k^\pm$ and, consequently, being related with the Painlev\'e IV equation \cite{Ber13,BF14,Bermudez14}. We will call them {\it Painlev\'e IV coherent states} (PIVCS). Moreover, due to the action of $l_k^\pm$ onto the eigenstates of the Hamiltonian, the Hilbert space $\mathcal{H}$ is naturally expressed as the direct sum of two subspaces: one of infinite dimension, related with the semi-infinite ladder arising from the original levels of the harmonic oscillator, and another one of finite dimension, associated with the new levels created by the SUSY transformation.

This work is organized as follows: in Section \ref{PHA and PIV}, the connection between an interesting class of quantum systems and the Painlev\'e IV equation will be established. The next section concerns with SUSY QM and the way to generate Painlev\'e IV Hamiltonians systems, along with their corresponding third-order ladder operators. In Section \ref{PIVCS}, several sets of coherent states for the aforementioned systems are generated, inside the finite and infinite subspaces. Our conclusions shall be presented in the last section.

\section{Polynomial Heisenberg algebras and Painlev\'e IV equation} \label{PHA and PIV}
The $m$th order polynomial Heisenberg algebras are defined by the following commutation relations:
\begin{equation}
\conm{H}{\mathcal{L}^\pm}=\pm \mathcal{L}^\pm,  \qquad \conm{\mathcal{L}^-}{\mathcal{L}^+}=N_{m+1}(H+1)-N_{m+1}(H)=P_m(H),\label{Capitulo4 PHA}
\end{equation}
where the operator $H$ is a Schr\"odinger Hamiltonian 
\begin{equation}
H=-\frac{1}{2}\frac{\text{d}^2}{\text{d}x^2}+V(x),  
\end{equation}
$N_{m+1}(H)$ is a polynomial of degree $m+1$ in $H$, which can be factorized as
\begin{equation}
N_{m+1}(H)=\mathcal{L}^+\mathcal{L}^-=\prod_{j=1}^{m+1}(H-\varepsilon_j),
\end{equation}
and thus  $P_m(H)$ is a polynomial of degree $m$ in $H$. Note that $\mathcal{L}^\pm$ are differential operators of $(m+1)$th order. In particular, for $m=0$ it is obtained that $N_1(H)= H-1/2$, $P_0(H)=\mathbb{1}$, $\mathcal{L}^+=a^+, \mathcal{L}^-=a^-$, recovering then the Heisenberg-Weyl algebra. 

Let us note that the algebras related to the Painlev\'e IV equation are of second order, arising for $m=2$ \cite{BF11a}. Indeed, in this case we have
\begin{equation}
\conm{\mathcal{L}^-}{\mathcal{L}^+}=P_2(H),
\end{equation}
with
\begin{equation}
N_3(H)=(H-\varepsilon_1)(H-\varepsilon_2)(H-\varepsilon_3). 
\end{equation}
The ladder operators $\mathcal{L}^\pm$, which are of third order, can be factorized as \cite{ACIN00,CFNN04}
\begin{equation}
\mathcal{L}^+=L_a^+L_b^+, \quad L_a^+= \frac{1}{\sqrt{2}}\left[-\frac{\text{d}}{\text{d}x}+f(x) \right], \quad L_b^+=\frac{1}{2} \left[\frac{\text{d}^2}{\text{d}x^2}+g(x)\frac{\text{d}}{\text{d}x}+h(x)  \right]. \label{Capitulo4 Lmas} 
\end{equation}
These operators satisfy the following intertwining relations: 
\begin{equation}
HL_a^+=L_a^+(H_a+1), \quad H_aL_b^+=L_b^+H \quad \Rightarrow \quad \conm{H}{\mathcal{L}^+}=\mathcal{L}^+,  
\end{equation}
where $H_a$ is an intermediate auxiliary Schr\"odinger Hamiltonian. Then, the functions $f,g,h,V_a$ and $V$ have to fulfill the following system of equations
\begin{subequations}\begin{align}
&-f'+f^2=2(V-\varepsilon_1),\\
&V_a=V+f'-1=V+g',\\
&\frac{g''}{2g}-\left( \frac{g'}{2g} \right)^2 -g'+\frac{g^2}{4}+\frac{(\varepsilon_2-\varepsilon_3)^2}{g^2}+\varepsilon_2+\varepsilon_3-2=2V,\\
&h=-\frac{g'}{2}+\frac{g^2}{2}-2V+\varepsilon_2 + \varepsilon_3-2.
\end{align}\end{subequations}
By decoupling this system, we obtain
\begin{subequations}\begin{align}
f & =g+x, \label{Capitulo4 f} \\
h & =+\frac{g'}{2}-\frac{g^2}{2}-2xg-x^2+a, \label{Capitulo4 h} \\
V & =\frac{x^2}{2}-\frac{g'}{2}+\frac{g^2}{2}+xg+\varepsilon_1-\frac{1}{2}, \label{Capitulo4 V}
\end{align}\end{subequations}
where $g(x)$ must satisfy
\begin{equation}
g''=\frac{g'^2}{2g}+\frac{3}{2}g^3+4xg^2+2(x^2-a)g+\frac{b}{g}, \label{Capitulo4 painleve}
\end{equation}
with $a=\varepsilon_2+\varepsilon_3-2\varepsilon_1-1, b=-2(\varepsilon_2-\varepsilon_3)^2$. This second order nonlinear differential equation is known as Painlev\'e IV equation. It is worth to note that the six Painlev\'e equations are second order nonlinear differential equations with the Painlev\'e properties, which in recent times have been studied in detail \cite{CM08,IKSY91,Lev92,BF11b}.

As can be seen, if one solution $g(x)$ of the Painlev\'e IV equation is obtained for certain values of $\varepsilon_1, \varepsilon_2,\varepsilon_3$, then the potential $V(x)$ as well as the corresponding ladder operators $L^\pm$ become completely determined. Moreover, the three extremal states, that are eigenstates of $H$ associated to $\varepsilon_i$ as well annihilated by $\mathcal{L}^-$, some of which could have physical interpretation, are given by
\begin{subequations}\begin{align}
\phi_{\varepsilon_1} (x)& \propto \exp \left(-\frac{x^2}{2} -\int g\, \text{d}x \right), \label{Capitulo4 psivarepsilon} \\
\phi_{\varepsilon_2} (x)& \propto \left(\frac{g'}{2g}-\frac{g}{2}-\frac{\varepsilon_2-\varepsilon_3}{g}-x \right) \exp\left[ \int \left(\frac{g'}{2g}+\frac{g}{2}-\frac{\varepsilon_2-\varepsilon_3}{g} \right)\, \text{d}x \right],  \\
\phi_{\varepsilon_3} (x)& \propto  \left(\frac{g'}{2g}-\frac{g}{2}+\frac{\varepsilon_2-\varepsilon_3}{g}-x \right) \exp\left[ \int \left(\frac{g'}{2g}+\frac{g}{2}+\frac{\varepsilon_2-\varepsilon_3}{g} \right)\, \text{d}x \right].   
\end{align}\end{subequations}

On the other hand, if we are able to identify a system ruled by third order differential ladder operators, it is possible to design a mechanism for obtaining solutions to the Painlev\'e IV equation. The key point of this procedure is to obtain the extremal states of our system; then, from the expression for the extremal state of Equation (\ref{Capitulo4 psivarepsilon}), it is straightforward to see that
\begin{equation}
g(x)=-x-\frac{\text{d}}{\text{d}x} \ln \left[\phi_{\varepsilon_1}(x) \right]. 
\end{equation}
Note that, by permuting cyclically the indices assigned to the extremal states we find three solutions to the Painlev\'e IV equation with different parameters $a, b$.

\section{Supersymmetric quantum mechanics and the harmonic oscillator}\label{SUSY} 

The SUSY QM is a technique which departs from a given solvable Hamiltonian $H_0$, with a complete set of orthogonal eigenvectors, and looks for another one $H_1$ whose eigenstates are yet to be obtained. The two Hamiltonians take the form
\begin{equation}
H_0=-\frac{1}{2}\frac{\text{d}^2}{\text{d}x^2}+V_0(x), \qquad H_1=-
\frac{1}{2}\frac{\text{d}^2}{\text{d}x^2}+V_1(x). \label{hamiltonianos a entrelazar}
\end{equation}  

In order to apply this technique, let us suppose the existence of a differential operator $A_1^\dag$ that intertwines the previous Hamiltonians in the way 
\begin{equation}
H_1 A_1^\dag = A_1^\dag H_0, \qquad A_1^\dag=\frac{1}{\sqrt{2}} \left[ -\frac{\text{d}}{\text{d}x}+ \frac{v'(x)}{v(x)}\right]. \label{entrelazamiento}
\end{equation}
Since $A_1^\dag$ is a first order differential operator, we refer to this case as 1-SUSY QM. It is also said that $V_0(x)$ and $V_1(x)$ are SUSY partner potentials. 

If we insert the explicit expressions for the Hamiltonians and the intertwining operator into Equation (\ref{entrelazamiento}), we find that $V_1(x)$ and $v(x)$ have to fulfill 
\begin{equation}
V_1(x)= V_0(x)- \frac{\text{d}^2}{\text{d}x^2} \ln v(x), \qquad - \frac{1}{2}v''(x)+V_0(x)v(x)=\epsilon v(x) ,
\end{equation}
where $\epsilon$ is an integration constant called {\it factorization energy}.

From the previous equations it can be seen that if we use a real solution $v(x)$ without zeros of the original stationary Schr\"odinger equation with factorization energy $\epsilon$, then the SUSY partner potential $V_1(x)$ is completely determined. Also, the intertwining relation (\ref{entrelazamiento}) ensures that if $\ket{n}$ is an eigenvector of $H_0$ with eigenvalue $E_n$, then $A_1^\dag \ket{n}$ will be an eigenstate of $H_1$ with the same eigenvalue. Note that the operators  $A_1^\dag $ and $A_1$ factorize the Hamiltonians $H_0$ and $H_1$ in the way
\begin{equation}
H_0 = A_1 A_1^\dag +\epsilon, \qquad H_1 = A_1^\dag A_1+\epsilon, \label{factorizacion}
\end{equation}
where $A_1 = (A_1^\dag)^\dag$. By evaluating the square of the norm of the vectors $A_1^\dag \ket{n}$ we have 
$$ 
||A_1^\dag \ket{n}||^2 = \left(\bra{n}A_1\right)\left( A_1^\dag \ket{n}\right)= \bra{n}\left(A_1 A_1^\dag \ket{n}\right)= E_n - \epsilon \geq 0 \quad \forall \ n,
$$ 
which implies that $ \epsilon \leq E_0$, where $E_0$ is the ground state energy of $H_0$. One could ask now if $\{A_1^\dag \ket{n}, \  n = 0,1,2, \dots \}$ is a complete orthogonal set. In order to answer this, let us assume the existence of a state $ \ket{\epsilon}$ which is orthogonal to every vector of the previous set, i.e.,
\begin{equation}
\bra{\epsilon}\left(A_1^\dag\ket{n}\right) = \left(\bra{\epsilon}A_1^\dag \right)\ket{n}= 0 \quad \forall \ n \qquad \Rightarrow \qquad A_1 \ket{\epsilon} = 0, 
\end{equation} 
since $\left\{ \ket{n}, \ n= 0,1,2,\dots \right\}$ is a complete orthogonal set. Let us choose $\phi_\epsilon(x) =\inner{x}{\epsilon}$ as the corresponding wavefunction, then the first-order differential equation $A_1 \phi_{\epsilon} = 0$ can be immediately solved to obtain
\begin{equation}
\phi_{\epsilon} (x)\propto  \frac{1}{v(x)}.
\end{equation}
Note that $\phi_{\epsilon}(x)$ satisfies:
\begin{equation}
H_1 \phi_{\epsilon} = \epsilon \phi_{\epsilon} .
\end{equation}
Thus, depending on the square integrability of this vector and the value of $\epsilon$, three possibilities arise:
\begin{itemize}
\item The vector $\ket{\epsilon}$, with $\epsilon < E_0$, belongs to the Hilbert space $\mathcal{H}$. Thus, $\{ \ket{\epsilon}, A^\dag_1 \ket{n}, \ n=0,1,2,\dots \}$ is a complete orthogonal set and, from Equations \eqref{entrelazamiento} and \eqref{factorizacion}, the spectrum of $H_1$ is given by Sp$[H_1]= \left\{ \epsilon, E_n, \, n=0,1,2,\dots \right\}$.

\item The state $\ket{\epsilon} \notin \mathcal{H}$, with $\epsilon < E_0$. In this case $\{ A^\dag_1 \ket{n}, \ n=0,1,2,\dots \}$ is a complete orthogonal set and thus Sp$[H_1]= \mathrm{Sp}[H_0]$.

\item The vector $\ket{\epsilon} \notin \mathcal{H}$ for $\epsilon = E_0$, then the set  $\{ A^\dag_1 \ket{n}, \ n=1,2,3,\dots\}$ is complete and thus Sp$[H_1]= \{ E_n, \ n=1,2,3, \dots \}$.
\end{itemize}

Summarizing, the new Hamiltonian $H_1$ will have a spectrum quite similar to the original one, differing perhaps in the ground state energy. In this work we will only focus on the first case, where a new level is inserted by the transformation. This technique can be iterated many times in order to obtain a Hamiltonian with a desired spectrum. 

\subsection{SUSY partners of the harmonic oscillator} \label{SUSY Partners}
Consider now a chain of $k+1$ Hamiltonians $H_j, j=0,1, \dots, k$, which are intertwined in the following way 
\begin{equation}
H_j A_j^\dag = A_j^\dag H_{j-1}, \qquad A_j^\dag = \frac{1}{\sqrt{2}} \left[ - \frac{\text{d}}{\text{d}x}+\frac{v'_j(x)}{v_j(x)}   \right], \qquad j=1, \dots k, \label{Capitulo4 Entrelazamiento}
\end{equation}
where
\begin{equation}
H_j=-\frac{1}{2}\frac{\text{d}^2}{\text{d}x^2}+V_j(x), \qquad j=0,1, \dots k,   
\end{equation}
i.e., $H_j$ and $H_{j-1}$ are SUSY partner Hamiltonians intertwined by the first order differential operator $A_j^\dag$. The potentials $V_j(x)$ and the transformation functions $v_j(x,\epsilon_{k-j})$ satisfy now
\begin{align}
H_{j-1}v_j=\epsilon_{k-j} v_j, \qquad  V_j(x)=V_{j-1}(x)- \frac{\text{d}^2}{\text{d}x^2}\ln v_j(x,\epsilon_{k-j}), \qquad j=1, \dots k.
\end{align}
In this way, if the potential $V_{j-1}(x)$ and transformation function $v_j(x,\epsilon_{k-j})$ are known, then the Hamiltonian $H_j$ is completely determined. Now, after composing the $k$ intertwining transformations induced by the operators $A_j^\dag, j=1,2, \dots.k$, and using Equation (\ref{Capitulo4 Entrelazamiento}), we get the following intertwining relation:   
\begin{equation}
H_k A_k^\dag \dots A_1^\dag = A_k^\dag \dots A_1^\dag H_0, \label{Capitulo4 entrelazamientok}
\end{equation}
i.e., the Hamiltonians $H_0$ and $H_k$ are intertwined by a $k$th order differential operator. To determine the Hamiltonian $H_k$, we need to know $k$ solutions $v_j(x,\epsilon_{k-j})$ of the stationary Schr\"odinger equations, one for each of the intermediate Hamiltonians. However, all of them can be obtained from solutions of the initial stationary Schr\"odinger equation. Indeed, if $u(x,E)$ is a solution of the equation 
\begin{equation}
H_0 u(x,E)=E u(x,E), \label{Capitulo 4 original}
\end{equation}
then, from the intertwining relation between $H_0$ and $H_1$ it is known that $v_2(x,E)\propto A_1^\dag u(x,E),$ $E\neq\epsilon$, will be a solution of $H_1 v_2(x,E)=E v_2(x,E)$. This procedure can be iterated to get the $k$ solutions $v_j(x,\epsilon_{k-j})$ from the corresponding ones of the initial Schr\"odinger equation $u(x,\epsilon)$. The potential $V_k(x)$ is given by  
\begin{equation}
V_k(x)=V_0(x)-\frac{\text{d}^2}{\text{d}x^2}\ln W\left[u(x,\epsilon_0),\dots,u(x,\epsilon_{k-1})\right], 
\end{equation}
where $W\left[ f_1,f_2, \dots, f_n \right] $ is the Wronskian of the functions $f_1,f_2, \dots, f_n$.  

In order to apply this technique to the harmonic oscillator, we need to know the general solution $u(x,\epsilon)$ of the Schr\"odinger equation for the potential $V_0(x)=x^2/2$ with an arbitrary factorization energy $\epsilon$, which is given by  
\begin{equation}
u(x,\epsilon)=e^{-x^2/2}\left[_1F_1\left(\frac{1-2\epsilon}{4},\frac{1}{2},x^2  \right)+2 \nu x \frac{\Gamma(\frac{3-2\epsilon}{4})}{\Gamma(\frac{1-2\epsilon}{4})}~_1F_1 \left(\frac{3-2\epsilon}{4},\frac{3}{2};x^2 \right) \right], \label{transformation function}
\end{equation}
where $\nu$ is a real arbitrary constant and $_1F_1(a,c,x)$ is the confluent hypergeometric function. If $\epsilon \leq E_0$, it is known that the solution will have no zeros for $|\nu|<1$ but it will have one node at one point of the real line for $|\nu|>1$. In order to generate a non singular potential $V_k(x)$ with $k$ new levels, it has to be chosen  $\epsilon_0< \epsilon_{1}< \dots <\epsilon_{k-1} < E_0$ with  $|\nu_{k-j}|<1$ for $j$ odd and $|\nu_{k-j}|>1$ for $j$ even, $j=1,2,\dots,k$. The new potential becomes now 
\begin{equation}
V_k(x)=\frac{x^2}{2}-\frac{\text{d}^2}{\text{d}x^2}\ln W\left[u(x,\epsilon_0),\dots,u(x,\epsilon_{k-1})\right], 
\end{equation}
and the spectrum of the  corresponding Hamiltonian $H_k$ will be 
\begin{equation}
\text{Sp}\left[H_k \right]=  \left\{ \epsilon_0, \epsilon_1,\dots,\epsilon_{k-1}, E_0,E_1,E_2,\dots\right\}.
\end{equation}

By denoting now $B_k^\dag=A_k^\dag \dots A_1^\dag$ and using the standard creation and annihilation operators for the harmonic oscillator $a^\pm$, it can be shown that
\begin{equation}
L_k^+ = B_k^\dag a^+ B_k, \qquad L_k^- = B_k^\dag a^- B_k,  \label{natural ladder}
\end{equation}
are $(2k+1)$-th order differential operators that obey the following commutation relations
\begin{equation}
\conm{H_k}{L_k^\pm}=\pm L_k^\pm, \label{Capitulo4 Conmutador}
\end{equation} 
i.e., they are the natural ladder operators for the Hamiltonian $H_k$.
	
Furthermore, from the intertwining relation (\ref{Capitulo4 entrelazamientok}) and the factorization of the intermediate Hamiltonians, $H_j=A^\dag_j A_j+\epsilon_{k-j},$ $H_{j-1}=A_jA_j^\dag+\epsilon_{k-j}$, it is obtained that
\begin{equation}
L_k^+L_k^-= \left(H_k-\frac{1}{2}\right) \prod_{j=0}^{k-1} (H_k-\epsilon_j)(H_k-\epsilon_j-1). \label{Capitulo4 PHAsusy}
\end{equation}
Comparing with Equation \eqref{Capitulo4 PHA}, it is seen that the set of operators $ \left\{ H_k, L_k^+, L_k^- \right\}$ generate a polynomial Heisenberg algebra of $2k$-th order, i.e., the natural ladder operators $L_k^\pm$ for the SUSY partners of the harmonic oscillator supply us with specific realizations of the general operators $\mathcal{L}^\pm$ generating the polynomial Heisenberg algebras.

\subsection{Painlev\'e IV Hamiltonian systems} \label{Hamiltonian systems}
The 1-SUSY partners of the harmonic oscillator can be used directly to find solutions to the Painlev\'e IV equation, since their natural ladder operators $L_1^+ \equiv B_1^\dag a^+ B_1, L_1^- \equiv B_1^\dag a^- B_1$ are of third order and thus they generate a second order polynomial Heisenberg algebra. Moreover, it has been recently found that some higher order SUSY partners of the harmonic oscillator also have third order ladder operators and, through them, new solutions of the Painlev\'e IV equation have been obtained \cite{BF11a}. This set of SUSY partners must satisfy the conditions contained in the following theorem.

\vspace{3mm}
\noindent {\bf Factorization Theorem}\\[5pt]
{\it Suppose that the $k$-th order SUSY partner $H_k$ of the harmonic oscillator Hamiltonian $H_0$ is generated by $k$ transformation functions $u(x,\epsilon_j), j=0,\dots,k-1,$ which are connected by the standard annihilation operator in the way:
\begin{equation}
u(x,\epsilon_{k-j-1}) = (a^-)^j u(x,\epsilon_{k-1}), \quad \epsilon_{k-j-1} = \epsilon_{k-1}-j,  
\end{equation} 
with $u(x,\epsilon_{k-1})$ being a nodeless solution of the stationary Schr\"odinger equation associated to $H_0$, given by Equation \eqref{transformation function} with $\epsilon_{k-1}<E_0=1/2$ and $|\nu_{k-1}|<1$. Therefore, the natural $(2k+1)$-th order ladder operator $L_k^+ \equiv B_k^\dag a^+ B_k$ of $H_k$ becomes factorized in the form 
\begin{equation}
L^+_k= P_{k-1}(H_k)l_k^+, 
\end{equation}
where $P_{k-1}(H_k)=(H_k-\epsilon_{1}) \dots (H_k-\epsilon_{k-1})$ is a polynomial of degree $k-1$ in $H_k$, $l^+_k$ is a third-order differential ladder operator such that $\left[H_k, l^+_k  \right]=l^+_k$, and
\begin{equation}
l^+_k l^-_k= \left(H_k-\frac{1}{2}\right)(H_k-\epsilon_0)(H_k-\epsilon_{k-1}-1). 
\end{equation} 
}

The proof of this theorem can be found in Reference \cite{BF11a}. It states that some Hamiltonians $H_k$, besides having their natural $(2k+1)$-th order differential ladder operators, also have third order ones. The corresponding spectrum contains now an equidistant ladder, with $k$ steps, below the ground state energy $E_0$ of $H_0$ plus the harmonic oscillator ladder. In addition, the transformation functions are no longer arbitrary: once the first one is chosen, all the others are automatically fixed by the theorem. As a result, the only parameters that remain free are now: the number of levels $k$ to be inserted, the energy gap $E_0-\epsilon_{k-1}$ between the two ladders (under the restriction $\epsilon_{k-1}<E_0$), and the real parameter $\nu_{k-1}$ of the transformation function $u(x,\epsilon_{k-1})$ (such that $|\nu_{k-1}|<1$).

\begin{figure}
\centering
\includegraphics[scale=0.5]{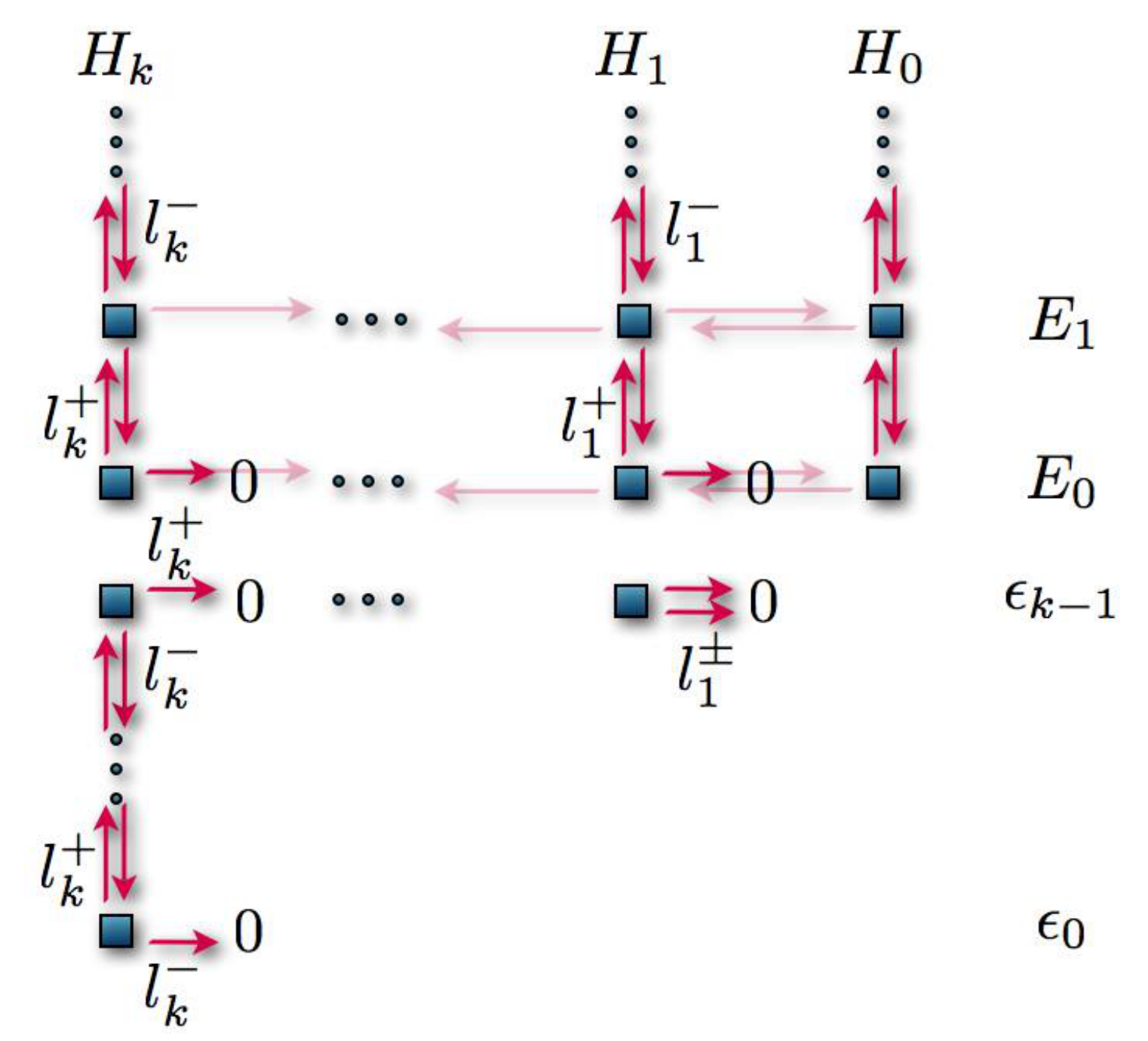}
\caption{Action of the operators $l_j^\pm$ on the eigenstates of the Hamiltonians $H_j, \ j=0,\dots,k$. We can see that $l_j^\pm$ allow the free displacement onto each independent ladder of $H_j$.}
\label{SUSY_partners}
\end{figure}

\section{Painlev\'e IV coherent states} \label{PIVCS}
Let us recall that the CS can be built up as eigenstates of the annihilation operator and also as the result of acting a certain displacement operator onto an extremal state.

In this section we will use the third-order ladder operators $l_k^\pm$ obtained in Section~\ref{Hamiltonian systems} to generate families of CS. Recall that these operators appear for very specific systems, which are ruled by second order PHA and consequently are directly connected with solutions to the Painlev\'e IV equation. To generate the CS, we will decompose the Hilbert space in two subspaces: one generated by the transformed eigenfunctions $\phi_n^{(k)}(x)$ (or the equivalent states $\ket{n^k}$ in Dirac notation) associated with the initial spectrum of the harmonic oscillator, which will be denoted as $\mathcal{H}_\text{iso}$; the other one is generated by the eigenfunctions $\phi_{\epsilon_j}^{(k)}$ (or the equivalent states $\ket{\epsilon_j^k}$ in Dirac notation) associated with the new energy levels, denoted as $\mathcal{H}_\text{new}$. 

The action of $l_k^{\pm}$ on the eigenstates $\ket{n^k} \in \mathcal{H}_{\text{iso}}$ of $H_k$ is given by
\begin{subequations}
\begin{align}
l_k^{-}\ket{n^k}&=\sqrt{(E_n-E_0)(E_n-\epsilon_0)(E_{n}-\epsilon_0-k)}\ket{n-1^{k}}, \label{cslkiso}\\
l_k^{+}\ket{n^k}&=\sqrt{(E_{n+1}-E_0)(E_{n+1}-\epsilon_0)(E_{n+1}-\epsilon_0-k)}\ket{n+1^{k}}. \label{cslka}
\end{align}\end{subequations}
On the other hand, on the remaining eigenstates $\ket{\epsilon_j^{k}} \in \mathcal{H}_{\text{new}}$ of $H_k$ we have
\begin{subequations}
\begin{align}
l_k^{-}\ket{\epsilon_j^k}&=\sqrt{(\epsilon_{j}-E_0)(\epsilon_j-\epsilon_0)(\epsilon_j-\epsilon_{0}-k)}\ket{\epsilon_{j-1}^k}, \label{cslknew}\\
l_k^{+}\ket{\epsilon_j^k}&=\sqrt{(\epsilon_{j+1}-E_0)(\epsilon_{j+1}-\epsilon_0)(\epsilon_{j+1}-\epsilon_{0}-k)}\ket{\epsilon_{j+1}^k},\label{cslk}
\end{align}\label{cslk2}\end{subequations}
where $E_0$ and $\epsilon_0$ are the lowest energy levels of $H_k$ in each of the two subspaces $\mathcal{H}_\text{iso}$ and $\mathcal{H}_\text{new}$, respectively. If we recall that $E_n=E_0+n$ and $\epsilon_j= \epsilon_0+j$, thus it is clear that we also will get the expected results for the two extremal states with $j=0$ and $j=k$ in $\mathcal{H}_\text{new}$. Note that the index $k$ in a generic vector $\ket{a^k}$ indicates the same label as the Hamiltonian $H_k$. Thus, for the eigenvectors the label $a$ will refer to the energy level, while for the CS we will have $a=z\in\mathbb{C}$ and we still will write the index $k$, in order to distinguish the new CS from those of the harmonic oscillator $\ket{z}$. 

It is worth to note that with this convention a confusion could appear when $z=\epsilon_j$ or when $z=n$; nevertheless, we believe that the context will make clear the specific situation we are dealing with. From Equations \eqref{cslkiso} and \eqref{cslknew} we can also see that $l_k^-$ annihilates the eigenstates $\ket{0^k}$ and $\ket{\epsilon_0^k}$, while from Equations~\eqref{cslka} and \eqref{cslk} it turns out that $l_k^+$ only annihilates $\ket{\epsilon_{k-1}^k}$ (see Figure~\ref{SUSY_partners}).

\subsection{Annihilation operator coherent states}
Now, for the Painlev\'e IV Hamiltonian systems we will generate the CS as eigenstates of the annihilation operator $l_k^-$, namely,
\begin{equation}
l_k^-\ket{z^k}=z\ket{z^k}. \label{csAOCSP4}
\end{equation}
Since in principle, we could generate independent CS in the two subspaces $\mathcal{H}_\text{iso}$ and $\mathcal{H}_\text{new}$, we are going to explore separately each of these two cases.

\subsubsection{PIVCS in the subspace $\mathcal{H}_{\text{iso}}$} \label{PIVCS AOCS}
In order to find the PIVCS $\ket{z^k_\text{iso}}$ in this subspace, we need to express this state as a linear combination of the eigenvectors $\left\{\ket{n^k},\, n=0,1,2, \dots \right\}$ of $H_k$, which form a complete orthonormal set in $\mathcal{H}_\text{iso}$. Therefore
\begin{equation}
\ket{z^k_\text{iso}}=\sum_{n=0}^\infty c_n\ket{n^k},
\end{equation}
where the constants $c_n$ are to be determined.

Applying $l_k^-$ on this expression, requiring that Equation~\eqref{csAOCSP4} is fulfilled and using Equation~\eqref{cslkiso} we finally obtain
\begin{equation}
\ket{z^k_\text{iso}}=c_0 \sum_{n=0}^{\infty}\frac{z^n}{\sqrt{n!}} \sqrt{\frac{\Gamma(E_0-\epsilon_0+1)\Gamma(E_0-\epsilon _0-k+1)}{\Gamma(E_0-\epsilon_0+1+n)\Gamma(E_0-\epsilon _0-k+1+n)}}\ket{n^k},\label{p4iso1}
\end{equation}
with an arbitrary constant $c_0$. Without lost of generality we can choose it as real positive, and by normalization of $\ket{z_\text{iso}^k}$ it turns out that
\begin{equation}
c_0= [{}_0F_2(E_0-\epsilon_0+1,E_0-\epsilon _0-k+1;|z|^2)]^{-1/2}, 
\end{equation} 
where ${}_pF_q$ is a generalized hypergeometric function defined as
\begin{equation}
_pF_q (a_1,\dots, a_p; b_1, \dots, b_q,x)\equiv \sum_{n=0}^\infty  \frac{(a_1)_n \dots (a_p)_n}{(b_1)_n \dots (b_q)_n} \frac{x^n}{n!}.\label{p4iso3}
\end{equation}
We also define an {\it auxiliary function} $c_0(a,b)$, which will be useful later on, as
\begin{equation}
c_0(a,b)=[{}_0F_2(E_0-\epsilon_0+1,E_0-\epsilon _0-k+1;a^{*}b)]^{-1/2}.
\end{equation}

Some mathematical and physical properties of these CS are the following: 

\begin{itemize}
\item \emph{Continuity of the labels.}
It is easy to check that if $z \rightarrow z'$ then $\norm{\ket{z'^k_\text{iso}}-\ket{z^k_\text{iso}}} \rightarrow 0$. Indeed
\begin{equation}
\norm{\ket{z'^k_\text{iso}}-\ket{z^k_\text{iso}}}^2=\inner{z'^k_\text{iso}-z^k_\text{iso}}{z'^k_\text{iso}-z^k_\text{iso}}=2\left[1-\text{Re}(\inner{z'^k_\text{iso}}{z^k_\text{iso}}) \right].  
\end{equation}
Let us write down the projection of two CS in the subspace, the so called {\it reproducing kernel}, using the auxiliary function $c_0(a,b)$, as
\begin{equation}
\inner{z'^k_\text{iso}}{z^k_\text{iso}}=\frac{c_0(z',z')c_0(z,z)}{c_0^2(z',z)}.
\end{equation}
Thus, in the limit $z' \rightarrow z$ it is found that $\ket{z'^k_\text{iso}} \rightarrow \ket{z^k_\text{iso}}$. In Figure~\ref{figproj2} we show the absolute value of this projection, $|\inner{z'^k_\text{iso}}{z_\text{iso}^k}|$, as function of $z$ for a fixed $z'$. Note that for the standard CS of the harmonic oscillator this plot would produce a Gaussian function, but in this case we find a certain deviation of that behavior.

\begin{figure}\centering
\includegraphics[scale=0.5]{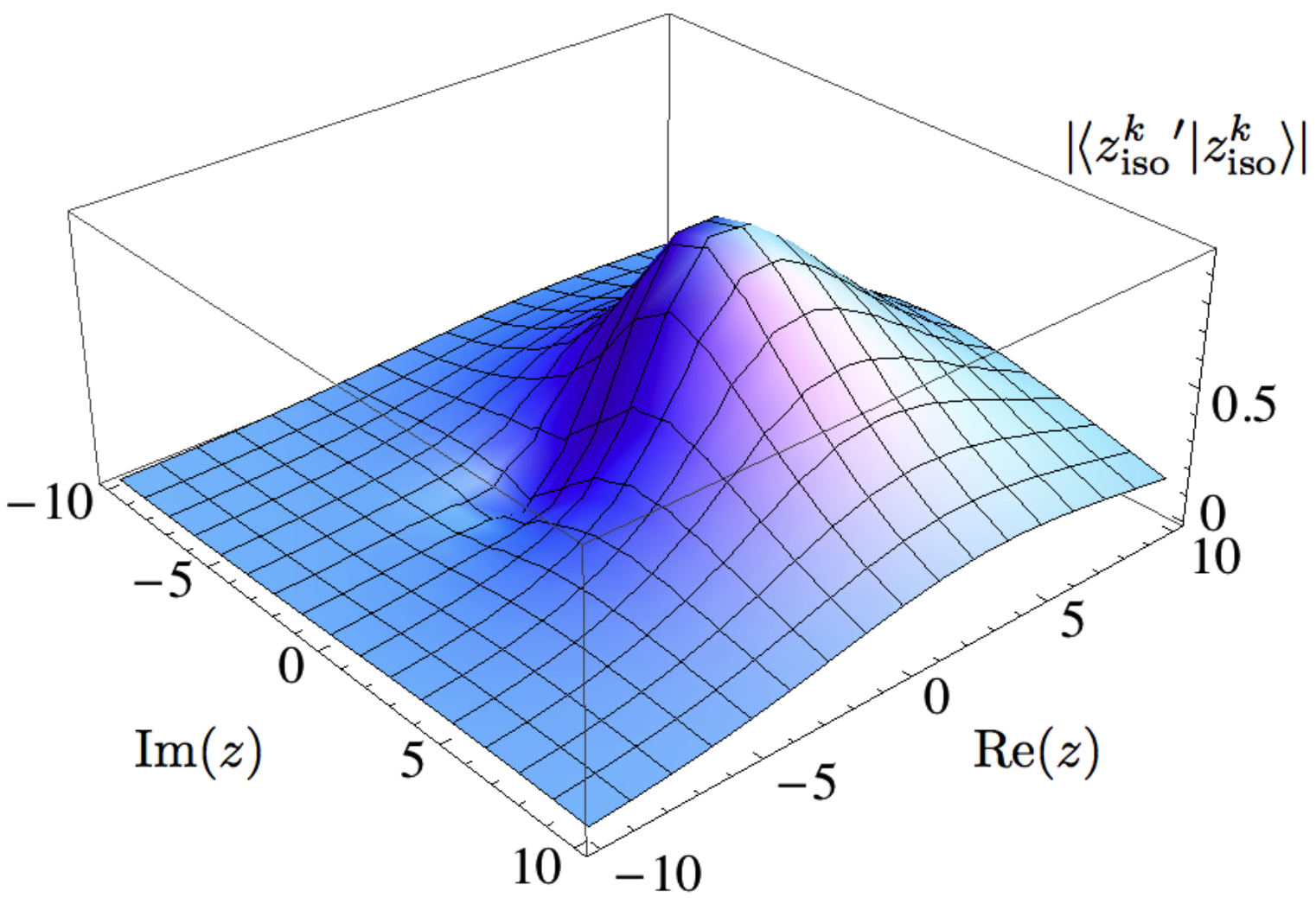}
\caption{\small{Absolute value of the projection of the CS $\ket{z_\text{iso}^k}$ onto another CS $\ket{z'^k_\text{iso}}$, both being eigenstates of the annihilation operator $l_k^-$, for $z'=5+i$, $\epsilon_0=-2$, and $k=2$.}}\label{figproj2}
\end{figure}

\item \emph{Resolution of the identity.}
We must look for a function $\mu_1(z)$ such that the following equation is fulfilled
\begin{equation}
\int\ket{z^k_\text{iso}}\bra{z^k_\text{iso}}\mu_1(z)\text{d}z=\mathbb{1}|_{\mathcal{H}_{\text{iso}}},\label{csidentity}
\end{equation}
i.e., these CS will satisfy the resolution of the identity operator in the subspace $\mathcal{H}_{\text{iso}}$. To accomplish this we propose
\begin{equation}
\mu_1 (z)= \frac{ f_1(|z|)}{\pi c_0^2(|z|)\Gamma(E_0-\epsilon_0+1) \Gamma(E_0-\epsilon_0-k+1)},
\end{equation}
insert this equation into \eqref{csidentity}, change $x=|z|^2=r^2$  and the summation index $n=s-1$. In the end we arrive to
\begin{equation}
\int_0^\infty x^{s-1} f_1(x)\text{d}x=\Gamma(E_0-\epsilon_0+s)\Gamma(E_0-\epsilon_0-k+s)\Gamma(s). \label{Medida AOCS}
\end{equation}
This means that $f_1(x)$ is the {\it inverse Mellin transform} of the right hand side of the last equation. Therefore $f_1(x)$ is given in terms of the {\it Meijer} $G$ {\it function}, defined as
\begin{equation}
\MeijerG{m}{n}{p}{q}{a_1,\ldots,a_p}{b_1,\ldots,b_q}{x}\equiv \mathcal{M}^{-1}\left[\frac{\prod\limits_{j=1}^{m}\Gamma(b_j+s)\prod\limits_{j=1}^n\Gamma(1-a_j-s)}{\prod\limits_{j=m+1}^{q}\Gamma(1-b_j-s)\prod\limits_{j=n+1}^p\Gamma(a_j+s)};x\right], \label{Gmeijer}  
\end{equation}
with $m=3$, $n=0$, $p=0$, $q=3$, $b_1=0$, $b_2=E_0-\epsilon_0$, $b_3=E_0-\epsilon_0-k$, i.e.,
\begin{equation}
f_1(r)=\MeijerG{3}{0}{0}{3}{0,E_0-\epsilon_0,E_0-\epsilon_0-k}{}{r^2}.  
\end{equation}
Notice that for $k=1$ we obtain the same results as Fern\'andez and Hussin \cite{FH99}. This is so because in that work the $(2k+1)$-th order differential ladder operators $L_k^\pm$ from Equation ~\eqref{natural ladder} are used to calculate a set of CS, while now we are using the third-order ladder operators $l_k^\pm$, and they coincide for $k=1$. However, for $k>1$ these states are different and completely new for the subspace $\mathcal{H}_{\text{iso}}$.

We still need to prove the positiveness of $\mu_1(z)$. To accomplish this, we follow the work of Sixdeniers and Penson \cite{SP00}, where a similar problem is solved using the {\it convolution property} for the inverse Mellin transform, also called {\it generalized Parseval formula}, which is given by
\begin{equation}
\mathcal{M}^{-1}[g^*(s)h^*(s);x] =\frac{1}{2\pi i}\int_{-i\infty}^{i\infty}g^*(s)h^*(s)x^{-s}\text{d}x =\int_0^{\infty}g(xt^{-1})h(t)t^{-1}\text{d}t, \label{parseval2}
\end{equation}
where $f^*(s)\equiv \mathcal{M}^{-1}[f(s);x]$. If we choose $h^*(s)=\Gamma(s)$, from the definition of the Gamma function we have
\begin{equation}
h(x)=\exp(-x) \label{Parseval AOCS h}
\end{equation}
which is a positive function for $x \geq 0$. 

Now, taking  $g^*(s)=\Gamma(E_0-\epsilon_0+s)\Gamma(E_0-\epsilon_0-k+s)$ and using Equation \eqref{Gmeijer} it is obtained 
\begin{equation}
g(x)=\MeijerG{2}{0}{0}{2}{E_0-\epsilon_0,E_0-\epsilon_0-k}{}{x}.  
\end{equation}
In addition, it turns out that \cite{Erd53}
\begin{eqnarray}
\MeijerG{2}{0}{0}{2}{a,b}{}{x}=2 x^{(a+b)/2}K_{a-b}(2x^{1/2}), \label{G a Bessel}
\end{eqnarray}
where $K_\nu(z)$ is a modified Bessel function of third kind. Using the integral representation of $K_\nu(z)$ given by \cite{AW01}
\begin{eqnarray}
K_\nu(z)=\frac{\pi^{1/2}}{\Gamma(\nu+1/2)}\left(\frac{z}{2}\right)^\nu \int_1^\infty e^{-z p}(p^2-1)^{\nu+1/2}\text{d}p, \quad \nu >-\frac{1}{2}, \quad -\frac{\pi}{2} < \text{arg} z < \frac{\pi}{2}, \label{Bessel integral} 
\end{eqnarray}
then $g(x)$ can be written as 
\begin{eqnarray}
g(x)=\frac{2 \pi^{1/2}}{\Gamma(k+1/2)}x^{E_0-\epsilon_0} \int_1^\infty e^{-2x^{1/2}p}(p^2-1)^{k-1/2} \text{d}p. \label{Parseval AOCS g}
\end{eqnarray}
It can be seen now that $g(x) \geq 0$ for $x \in [0,\infty)$. Then, inserting Equations \eqref{Parseval AOCS h} and \eqref{Parseval AOCS g} in \eqref{parseval2}, the positiveness of $f_1(r)$ of Equation \eqref{Medida AOCS} is guaranteed and hence the positiveness of the measure $\mu_1(z)$. 

\item \emph{Temporal stability.}
If we apply the evolution operator to the CS in the subspace $\ket{z^k_{\text{iso}}}$ we obtain
\begin{align}
\hspace{-2mm}U(t)\ket{z^k_{\text{iso}}}=&c_0\exp(-iH_kt)\sum_{n=0}^\infty\frac{z^n}{\sqrt{n!}}\sqrt{\frac{\Gamma(E_0-\epsilon_0+1)\Gamma(E_0-\epsilon _0-k+1)}{\Gamma(E_0-\epsilon_0+1+n)\Gamma(E_0-\epsilon _0-k+1+n)}}\ket{n^k}   \nonumber \\
=& \exp(-iE_0t)\ket{z\exp(-it)^k_\text{iso}}\equiv \exp(-iE_0t)\ket{z^k_\text{iso}(t)}.
\end{align}
This means that, up to a global phase factor, a CS evolves into another CS in the same subspace.

\item \emph{Mean energy value.}
The mean value of the energy in a CS can be directly calculated using the explicit expression of the CS given by Equation \eqref{p4iso1}: 
\begin{align}
\esp{H_k}_{z_\text{iso}}&=\bra{z_\text{iso}^k}H_k\ket{z_\text{iso}^k}  \nonumber \\ &= \frac{1}{2}+ \frac{|z|^2}{(E_0-\epsilon_0+1)(E_0-\epsilon_0-k+1)} \frac{{}_0F_2(E_0-\epsilon_0+2,E_0-\epsilon _0-k+2;|z|^2)      }{{}_0F_2(E_0-\epsilon_0+1,E_0-\epsilon _0-k+1;|z|^2)}.
\end{align}

\item \emph{State probability.}
It is also useful to calculate the probability $p_n(z)$ that an energy measurement for the system being in a CS $\ket{z^k_{\text{iso}}}$ gives the value $E_n$. This probability $p_n(z)$ turns out to be:
\begin{equation}
p_n(z) =|\inner{n^k}{z^k_\text{iso}}|^2=c_0^2\frac{|z|^{2n}}{n!}\frac{\Gamma(E_0-\epsilon_0+1)\Gamma(E_0-\epsilon _0-k+1)}{\Gamma(E_0-\epsilon_0+1+n)\Gamma(E_0-\epsilon _0-k+1+n)}.
\end{equation}\label{prob1}
\end{itemize}

Finally, all these properties mean that the states $\ket{z_\text{iso}^k}$ constitute an appropriate set of CS in the subspace $\mathcal{H}_{\text{iso}}$.

\subsubsection{PIVCS in the subspace $\mathcal{H}_{\text{new}}$}
The subspace $\mathcal{H}_{\text{new}}$ is $k$-dimensional; therefore, the operator $l_k^-$ can be represented by a $k \times k$ matrix with elements given by
\begin{equation}
(l_k^-)_{mn}=\inner{\epsilon_m^k}{l_k^-|\epsilon_n^k}.
\end{equation}
Then, from Equation~\eqref{cslknew} we see that its only non null elements are in the so called {\it superdiagonal}, i.e., directly above the main diagonal. Furthermore, it is straightforward to check that this matrix is {\it nilpotent}, with its $k$th-power being the {\it zero matrix}.

Now, multiplying the eigenvalue equation $(l_k^-) {\bf x} = z {\bf x}$, by $(l_k^-)^{k-1}$ we obtain
\begin{equation}
(l_k^-)^k {\bf x} = z^k {\bf x} = {\bf 0} \qquad \Rightarrow \qquad z=0,  
\end{equation}
i.e., the only possible eigenvalue for the matrix $(l_k^-)$ is $z=0$. The same turns out to be valid for $l_k^-$ and then, its only eigenvector in $\mathcal{H}_{\text{new}}$ is $\ket{\epsilon_0^k}$. Therefore, through this definition we cannot generate a family of CS in the subspace $\mathcal{H}_{\text{new}}$ that satisfies the resolution of the identity operator in this subspace. This is due to the finite dimension of $\mathcal{H}_{\text{new}}$.

\subsection{Displacement operator coherent states}
The CS defined as displaced versions of the ground state are not simple to generate for the $k$-SUSY partner Hamiltonians $H_k$ of the harmonic oscillator since the commutator of $l_k^-$ and $l_k^+$ is no longer the identity operator but a second degree polynomial in $H_k$. Therefore, if we change $a^- \rightarrow l_k^-$ and $a^+ \rightarrow l_k^+$ in the displacement operator for the harmonic oscillator, it turns out that
\begin{equation}
\widetilde{D}(z) = \exp\left(z l_k^+ - z^*l_k^- \right) \neq \exp\left(-\frac{1}{2}|z|^2 \right) \exp \left(z l_k^+ \right) \exp  \left( -z^*l_k^- \right),
\end{equation}
i.e., now it is not so simple to separate $\widetilde{D}(z)$ into exponentials. For that reason, we decided to propose instead an operator already factorized from the very beginning, i.e., the right hand side of this last expression is going to be taken as the displacement operator for the new systems,
\begin{equation}\label{displacementoperator}
D(z)=\exp\left(-\frac{1}{2}|z|^2 \right) \exp \left(z l_k^+ \right) \exp  \left( -z^*l_k^- \right),
\end{equation}
although it is not a unitary operator.

Now, let us recall that for the harmonic oscillator, the ground state is annihilated by $a^-$. A generalization of this procedure consists in using not the ground but an extremal state, i.e., a non-trivial eigenstate of $H_k$ belonging as well to the kernel of the annihilation operator $l_k^-$. There are two such extremal states for $H_k$: the state $\ket{0^k}$ in the subspace $\mathcal{H}_\text{iso}$ and $\ket{\epsilon^k_0}$ in $\mathcal{H}_\text{new}$. Once again, let us explore separately each of these two cases.

\subsubsection{PIVCS in the subspace $\mathcal{H}_{\text{iso}}$}
We apply the previously defined displacement operator $D(z)$ onto the extremal state $\ket{0^k}\in\mathcal{H}_{\text{iso}}$, adding also a normalization constant $C_z$ for convenience,
\begin{equation}
\ket{z^k_{\text{iso}}}= C_z D(z) \ket{0^k}= C_z \exp\left(-\frac{1}{2}|z|^2 \right)\sum_{n=0}^\infty \frac{(zl_k^+)^n}{n!}\ket{0^k}. \end{equation}
After several calculations we obtain
\begin{equation}
\ket{z^k_{\text{iso}}}=C_z \exp\left(-\frac{1}{2}|z|^2 \right) \sum_{n=0}^\infty \frac{z^n}{\sqrt{n!}}\left[\prod_{m=1}^n\sqrt{(m+E_0-\epsilon_0)(m+E_0-\epsilon_0-k)}\right]\ket{n^k}.  
\end{equation}
At first sight one could think that this is a right set of CS in this subspace for $z\in {\mathbb C}$. Nevertheless, if we analyze its normalization it is found that
\begin{align}
\inner{z_\text{iso}^k}{z_\text{iso}^k}=&|C_z|^2 \exp\left(-|z|^2 \right)\sum_{n=0}^\infty\frac{|z|^{2n}}{n!}\prod_{m=1}^n (m+E_0-\epsilon_0)(m+E_0-\epsilon_0-k)  \nonumber \\
=&|C_z|^2 \exp\left(-|z|^2 \right)\, {}_2F_0(E_0+1-\epsilon_0,E_0+1-\epsilon_0-k;|z|^2).
\end{align}
The fact that it is expressed in terms of the generalized hypergeometric function ${}_2F_0(E_0+1-\epsilon_0,E_0+1-\epsilon_0-k;|z|^2)$ indicates that the norm can be made equal to $1$ only when $z=0$, but it diverges for all $\mathbb{C} \ni z\neq 0$ \cite{Erd53}. Therefore, the only square-integrable CS appearing when we apply this displacement operator onto the extremal state $\ket{0^k}$ in $\mathcal{H}_{\text{iso}}$ is precisely $\ket{0^k}$. For $z\neq 0$ we obtain an expression that does not correspond to any vector in the Hilbert space of the system.

\subsubsection{PIVCS in the subspace $\mathcal{H}_{\text{new}}$}
Let us apply now the displacement operator $D(z)$ onto the extremal state $\ket{\epsilon_0^k}\in \mathcal{H}_{\text{new}}$, which is also annihilated by $l_k^-$. This leads us to
\begin{equation}
\ket{z_\text{new}^k}= N_z' \exp\left(-\frac{|z|^2}{2}\right) \left[\sum_{j=0}^{k-1}\left( \prod_{i=1}^j \sqrt{(E_0-\epsilon_0-i)(k-i)}\right) \frac{z^j}{\sqrt{j!}} \ket{\epsilon_j^k}  \right]. \label{csdisp1}
\end{equation}
We use the factor $\exp(-|z|^2/2)$ and the constant $N_z'$ to define $N_z=N_z' \exp(-|z|^2/2)$; we also employ the definition of the Pochhammer symbols
\begin{equation}
(x)_n\equiv x(x+1)(x+2)\dots(x+n-1)=\frac{\Gamma(x+n)}{\Gamma(x)},
\end{equation}
to rewrite
\begin{subequations}
\begin{align}
\prod_{i=1}^j(k-i)&=\frac{\Gamma(k)}{\Gamma(k-j)} = (k-j)_j,\\
\prod_{i=1}^j(E_0-\epsilon_0-i)&=\frac{\Gamma(E_0-\epsilon_0)}{\Gamma(E_0-\epsilon_0-j)}=(E_0-\epsilon_0-j)_j.
\end{align}
\end{subequations}
Then we have
\begin{equation}
\ket{z_\text{new}^k}= N_z \left[\sum_{j=0}^{k-1}\sqrt{(E_0-\epsilon_0-j)_j (k-j)_j} \frac{z^j}{\sqrt{j!}} \ket{\epsilon_j^k}  \right].\label{csdisp3}
\end{equation}

In this case we do not have any problem with the normalization, because the involved sum is finite. Without lost of generality we can choose $N_z$ to be real positive such that $\langle z_\text{new}^k \ket{z_\text{new}^k} = 1$ and hence
\begin{equation}
N_z=\left[\sum_{j=0}^{k-1}\frac{|z|^{2j}}{j!}(E_0-\epsilon_0-j)_j (k-j)_j\right]^{-1/2}.\label{csdisp2}
\end{equation}

Some properties of the set $\left\{ \ket{z_\text{new}^k} \right\}|_{z \in  \mathbb{C}}$ are the following:
\begin{itemize}
\item \emph{Continuity of the labels.}
The proof is similar as for the annihilation operator CS
\begin{equation}\label{normacuadrada}
\norm{\ket{z'^k_\text{new}}-\ket{z^k_\text{new}}}^2=\inner{z'^k_\text{new}-z^k_\text{new}}{z'^k_\text{new}-z^k_\text{new}}=2\left[1-\text{Re}(\inner{z'^k_\text{new}}{z^k_\text{new}}) \right].  
\end{equation}
The normalization factor $N_z$ of Equation \eqref{csdisp2} suggests to define a more general function
\begin{equation}
N(a,b)=\left[\sum_{j=0}^{k-1}\frac{(a^*b)^{j}}{j!}(E_0-\epsilon_0-j)_j (k-j)_j \right]^{-1/2}.
\end{equation}
Using this definition we obtain a simple expression for the inner product in \eqref{normacuadrada}:
\begin{equation}
\inner{z'^k_\text{new}}{z^k_\text{new}}=\frac{N(z',z')N(z,z)}{N^2(z',z)}.
\end{equation}
This means that in the limit $z' \rightarrow z$, we get $\ket{z'^k_\text{new}} \rightarrow \ket{z^k_\text{new}}$. 

\item \emph{Resolution of the identity.}
In this case we should show that
\begin{equation}
\int\ket{z_\text{new}^k}\bra{z_\text{new}^k}\mu_2 (z)\text{d}z=\mathbb{1}|_{\mathcal{H}_\text{new}}.
\end{equation}
By plugging the expression for the CS of Equation \eqref{csdisp3}, it turns out that
\begin{equation}
\mathbb{1}|_{\mathcal{H}_\text{new}}=2\pi\sum_{j=0}^{k-1}\frac{\ket{\epsilon_j^k}\bra{\epsilon_j^k}}{j!}\frac{\Gamma(E_0-\epsilon_0)}{\Gamma(E_0-\epsilon_0-j)}\frac{\Gamma(k)}{\Gamma(k-j)}\int_0^\infty N_z^2 r^{2j+1}\mu_2 (r)\text{d}r,
\end{equation}
where we have used polar coordinates, assumed that $\mu_2 (z) = \mu_2 (r)$, and integrate the angle variable. Then we propose that 
\begin{equation}
\mu_2(r)=\frac{f_2(r)}{\pi N_z^2\Gamma(E_0-\epsilon_0)\Gamma(k)}.
\end{equation}
We can change now $x=r^2$ and $s=j+1$ to obtain the condition on $f_2(r)$ as
\begin{equation}
\int_0^\infty x^{s-1}f_2(x)\text{d}x=\Gamma(1+k-s)\Gamma(1+E_0-\epsilon_0-s)\Gamma(s).
\end{equation}
It turns out that $f_2(r)$ is also a Meijer $G$ function,
\begin{equation}
f_2(r)=\MeijerG{1}{2}{2}{1}{-k,\epsilon_0-E_0}{0}{r^2}. \label{G DOCS}
\end{equation}

Once again, we need to prove the positiveness of $\mu_2(r)$. We will proceed now as in Section \ref{PIVCS AOCS} with the generalized Parseval formula \eqref{parseval2} in order to express \eqref{G DOCS} in integral form. Choosing again $h^*(s)=\Gamma(s)\Rightarrow h(x)=\exp(-x)$, which is a positive function for $x \in [0,\infty)$. Taking now $g^*(s)=\Gamma(k+1-s)\Gamma(E_0-\epsilon_0+1-s)$ we have
\begin{equation}
g(x)=\MeijerG{0}{2}{2}{0}{-k,\epsilon_0-E_0}{}{x}.  
\end{equation}
Using the following property of the Meijer $G$ function \cite{Erd53}
\begin{eqnarray}
\MeijerG{m}{n}{p}{q}{a_r}{b_s}{x^{-1}}=\MeijerG{n}{m}{q}{p}{1-b_s}{1-a_r}{x},
\end{eqnarray}
as well as the identity \eqref{G a Bessel}, the fact that $K_\nu(z)= K_{-\nu}(z)$ and the integral representation of the Bessel function $K_\nu(z)$ given by Equation \eqref{Bessel integral}, we can express $g(x)$
as
\begin{eqnarray}
g(x)=\frac{2 \pi^{1/2}}{\Gamma(E_0 -\epsilon_0-k+1/2)}x^{-(E_0 -\epsilon_0-2k+1)} \int_1^\infty e^{-2x^{-1/2}p}\left(p^2-1 \right)^{E_0-\epsilon_0-k-1/2}\text{d}p, \label{g docs 1}
\end{eqnarray}
which is valid for $E_0-\epsilon-k > -1/2$. The last condition is not always fulfilled since in the system under consideration we have  $E_0-\epsilon_0-k >-1$, which means that the $k$ inserted levels are below the ground state $E_0$ of the harmonic oscillator. For the interval $-1< E_0-\epsilon-k < -1/2$ the appropriate expression is  
\begin{eqnarray}
g(x)=\frac{2 \pi^{1/2}}{\Gamma(\epsilon_0+k-E_0+1/2)}x^{k+1} \int_1^\infty e^{-2x^{-1/2}p}\left(p^2-1 \right)^{-(E_0-\epsilon_0-k+1/2)}\text{d}p.\label{g docs 2}
\end{eqnarray}
It can be seen in both cases that $g(x) \geq 0$ for $x\in(0,\infty)$ (Equations \eqref{g docs 1} and \eqref{g docs 2}). Using the generalized Parseval formula, this result ensures that $\mu_2(r)$ is a positive definite measure.

\item \emph{Temporal stability.}
If the evolution operator is applied to a CS in the subspace $\mathcal{H}_{\text{new}}$ we obtain
\begin{align}
U(t)\ket{z^k_{\text{new}}}&=\exp(-iH_kt)N_z \sum_{j=0}^{k-1}\sqrt{(E_0-\epsilon_0-j)_j(k-j)_j}\frac{z^j}{\sqrt{j!}}\ket{\epsilon_j^k}  \nonumber \\
&=\exp(-i\epsilon_0 t)\ket{z\exp(-it)^k_{\text{new}}}\equiv \exp(-i\epsilon_0t)\ket{z^k_{\text{new}}(t)}.
\end{align}
We can see that, up to a global phase factor, one of these CS evolves always into another CS in the same subspace.

\item \emph{Mean energy value.}
In order to evaluate the mean value of the energy we use the explicit expression of the CS of Equation \eqref{csdisp3}. The result is the following:
\begin{equation}
\esp{H_k}_{z_\text{new}}=\bra{z_\text{new}^k}H_k\ket{z_\text{new}^k}
=\epsilon_0 + N_z^2 \left[\sum_{j=0}^{k-1}j(E_0-\epsilon_0-j)_j(k-j)_j \frac{|z|^{2j}}{j!} \right].
\end{equation}

\item \emph{State probability.}
For a system being in a CS $\ket{z_\text{new}^k}$, the probability $p_j(z)$ to obtain the energy $\epsilon_j$ is now given by 
\begin{equation}
p_j(z)=|\inner{\epsilon_j^k}{z^k_\text{new}}|^2 = N_z^2 (E_0-\epsilon_0-j)_j (k-j)_j \frac{|z|^{2j}}{j!}.
\end{equation}

\end{itemize}

\subsection{Linearized displacement operator coherent states}
We have seen that the definition of CS as eigenstates of the annihilation operator $l_k^-$ works appropriately only for $\mathcal{H}_{\text{iso}}$ and the one associated to the displacement operator $D(z)$ only for $\mathcal{H}_{\text{new}}$, i.e., until now no definition allows us to obtain sets of CS in the two subspaces $\mathcal{H}_{\text{iso}}$ and $\mathcal{H}_{\text{new}}$ simultaneously when using the third-order ladder operators $l_k^\pm$. Nevertheless, we still have the alternative to {\it linearize} $l_k^\pm$, i.e., to define some new ladder operators as
\begin{subequations}\begin{align}
\ell_k^{+}&\equiv \sigma(H_k)l_k^{+},   \\
\ell_k^{-}&\equiv \sigma(H_k+1)l_k^{-},\label{csellk}
\end{align}\label{p4linell}\end{subequations}
\hspace{-1mm}where
\begin{equation}
\sigma(H_k)=[(H_k-\epsilon_0)(H_k-\epsilon_0-k)]^{-1/2},  
\end{equation}
and by convention we take the positive square root. The infinite-order differential ladder operators $\ell_k^\pm$ can be alternatively defined through their action onto the basis of $\mathcal{H}_{\text{iso}}$ and $\mathcal{H}_\text{new}$.

Note that it would seem more natural to define $\ell_k^{-}$ as $(\ell_k^{+})^\dag$, but in such a case we would not have a well defined action of $\ell_k^{-}$ onto $\ket{\epsilon_0^k}$, i.e., in general the action of the two alternative definitions is different. However, the ladder operators of Equations \eqref{p4linell} act on the eigenvectors of the Hamiltonian $H_k$ in a strongly simplified way, which justifies its definition. This linearization process has been applied previously to the general SUSY partners of the harmonic oscillator in order to obtain families of CS using different annihilation operators \cite{FHN94,FNR95,FH99, FHR07}.

\subsubsection{PIVCS in the subspace $\mathcal{H}_{\text{iso}}$}
The new ladder operators $\ell_k^\pm$ act on the eigenvectors of $H_k$ in the subspace $\mathcal{H}_{\text{iso}}$ as follows:
\begin{subequations}
\begin{align}
\ell_k^{-}\ket{n^k} & =\sqrt{n}\ket{n-1^k},\\
\ell_k^{+}\ket{n^k} & =\sqrt{n+1}\ket{n+1^k}.  
\end{align}
\end{subequations}
Let us define now the analogue number operator in $\mathcal{H}_{\text{iso}}$ as $N\equiv \ell_k^{+}\ell_k^{-}$, given that $N\ket{n^k}=n\ket{n^k}$. Furthermore, we can easily show that the operators $\{\ell_k^{+},\ell_k^{-},H_k, \mathbb{1}\}$ obey the following commutation rules inside $\mathcal{H}_{\text{iso}}$:
\begin{equation}
[\ell_k^{-},\ell_k^{+}]=\mathbb{1},\quad [H_k,\ell_k^{\pm}]=\pm\ell_k^{\pm}.\label{csalgebra}
\end{equation}
Equations~\eqref{csalgebra} mean that the linearized ladder operators satisfy the Heisenberg-Weyl algebra on $\mathcal{H}_\text{iso}$.

In order to generate a set of linearized CS, let us define now an analogue of the displacement operator as
\begin{equation}
\mathcal{D}(z)= \exp\left(-\frac{1}{2}|z|^2 \right) \exp \left(z \ell_k^+ \right) \exp  \left( -z^*\ell_k^- \right),\label{lineD}
\end{equation}
i.e., similar to Equation \eqref{displacementoperator} but with the linearized ladder operators $\ell_k^\pm$ placed instead of $l_k^\pm$. Then, the CS turn out to be
\begin{equation}
\ket{z^k_{\text{iso}}}= \mathcal{D}(z)\ket{0^k}=\exp\left(-\frac{|z|^2}{2}\right)\sum_{n=0}^{\infty}\frac{z^n}{\sqrt{n!}}\ket{n^k},\label{linstates}
\end{equation} 
i.e., in the subspace $\mathcal{H}_{\text{iso}}$ the linearized ladder operators lead to an expression similar to the CS of the harmonic oscillator (see Equation \eqref{standard CS}). The difference rely in the states that are involved: for the harmonic oscillator they are the eigenstates of $H_0$, while in this case the involved states are the eigenstates of $H_k$ in $\mathcal{H}_{\text{iso}}$. Note that the states of Equation \eqref{linstates} are already normalized.

Let us analyze some properties of this new set of CS.

\begin{itemize}
\item \emph{Continuity of the labels.}
The proof that $\ket{z'^k_{\text{iso}}} \rightarrow \ket{z^k_{\text{iso}}}$ when $z' \rightarrow z$ is similar to the one for the CS of the annihilation operator in $\mathcal{H}_{\text{iso}}$.

\item \emph{Resolution of the identity.}
To prove the identity resolution the same procedure as for the harmonic oscillator is followed to obtain \cite{Coh77}
\begin{equation}
\frac{1}{\pi}\iint \ket{z^k_{\text{iso}}} \bra{z^k_{\text{iso}}}\, \text{d}\, \text{Re}(z)\, \text{d}\, \text{Im}(z)= \mathbb{1}|_{\mathcal{H}_\text{iso}}.
\end{equation}
In this way we ensure that any vector belonging to $\mathcal{H}_{\text{iso}}$ can be expressed in terms of these CS.

\item \emph{Temporal stability.}
When the evolution operator is applied to the CS $\ket{z^k_{\text{iso}}}$ we obtain
\begin{align}
U(t)\ket{z^k_{\text{iso}}}&=\exp(-iH_kt)\exp\left(-\frac{|z|^2}{2}\right)\sum_{n=0}^\infty\frac{z^n}{\sqrt{n!}}\ket{n^k} \nonumber \\
&=\exp(-iE_0t)\exp\left(-\frac{|z|^2}{2}\right)\sum_{n=0}^\infty\frac{[z\exp(-it)]^n}{\sqrt{n!}}\ket{n^k} \nonumber \\
&=\exp(-iE_0t)\ket{z\exp(-it)^k_\text{iso}}\equiv \exp(-iE_0t)\ket{z^k_\text{iso}(t)}.
\end{align}
This means that, up to a global phase factor, any CS $\ket{z^k_{\text{iso}}}$ evolves into another CS in the same subspace.

\item \emph{Mean energy value.}
When the operator $\ell_k^-$ is applied on a coherent state $\ket{z_\text{iso}^k}$ it is obtained 
\begin{eqnarray}
\ell_k^- \ket{z_\text{iso}^k} = z \ket{z_\text{iso}^k},
\end{eqnarray}
i.e., the CS arising when we act the displacement operator of equation \eqref{lineD} onto the extremal state $\ket{0^k}$ also can be generated as eigenstates of the linearized annihilation operator. This result is used to calculate the mean energy value in a CS as follows:
\begin{align}
\esp{H_k}_{z_\text{iso}}&=\bra{z_\text{iso}^k}H_k\ket{z_\text{iso}^k} = \bra{z_\text{iso}^k}(\ell^+_k \ell^-_k + E_0)\ket{z_\text{iso}^k}=|z|^2+E_0.   \end{align}

\item \emph{State probability.}
If the system is in a CS $\ket{z_{\text{iso}}^k}$ and we perform an energy measurement, then the probability $p_n(z)$ of getting the value $n+1/2$ is given by
\begin{equation}
p_n(z)=|\inner{n^k}{z_{\text{iso}}^k}|^2=e^{-|z|^2} \frac{|z|^{2n}}{n!},\label{prob2}
\end{equation}
which is a Poisson distribution with mean value at $|z|^2$. 

An example of the modulus squared of the wavefunction associated to a CS $\ket{z_\text{iso}^k}$ is shown in Figure 3.
\end{itemize}

\begin{figure}[ht]
\centering \includegraphics[width=12cm]{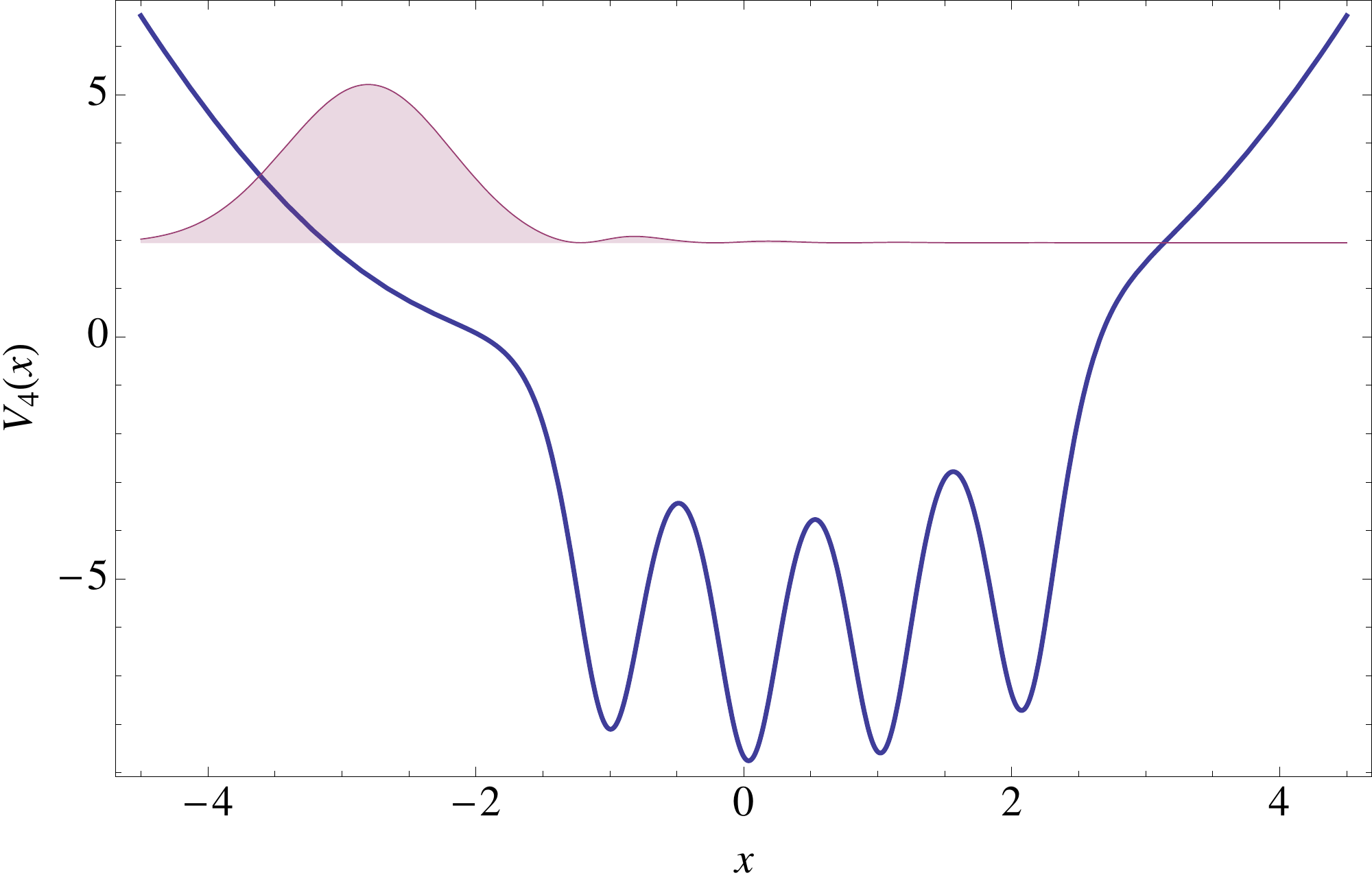}
\caption{The modulus squared of the wavefunction associated to a coherent state $\ket{z_\text{iso}^k}$ in $\mathcal{H}_{\text{iso}}$ with $z=1.2 e^{-2.78 i}$ and mean energy value of $1.94$ for a Painlev\'e IV Hamiltonian system obtained with $k=4$, $\epsilon_3=-2.8$, and $\nu_3=-0.9$.}
\label{Figure LinearizedCSiso}
\end{figure}

\subsubsection{PIVCS in the subspace $\mathcal{H}_{\text{new}}$}
The action of the linearized ladder operators $\ell_k^\pm$ onto the basis of $\mathcal{H}_{\text{new}}$, $\left\{ \ket{\epsilon_j^k} , j=0,\dots,\right.$ $\left.k-1 \right\}$, is given by
\begin{subequations}\begin{align}
\ell_k^{+}\ket{\epsilon_j^k}&=(1-\delta_{j,k-1})\sqrt{\epsilon_{j+1}-E_0}\ket{\epsilon_{j+1}^k} , \\ \ell_k^{-}\ket{\epsilon_j^k}&=(1-\delta_{j,0})\sqrt{\epsilon_{j}-E_0}\ket{\epsilon_{j-1}^k},
\end{align}\end{subequations}
which comes from the definition of $\ell_k^\pm$ in Equations \eqref{p4linell}. It turns out that $\ell_k^-$ annihilates the eigenstate $\ket{\epsilon_0^k}$ while $\ell_k^+$ annihilates $\ket{\epsilon_{k-1}^k}$. Now, the commutator $\conm{\ell_k^-}{\ell_k^+}$ acts on the same basis in the way:
\begin{subequations}\begin{align}
[\ell_k^{-},\ell_k^{+}]\ket{\epsilon_0^k}&=(\epsilon_0+1-E_0)\ket{\epsilon_0^k},   \\
[\ell_k^{-},\ell_k^{+}]\ket{\epsilon_j^k}&=\ket{\epsilon_j^k},\quad \quad j=1,2,\dots ,k-2,   \\
[\ell_k^{-},\ell_k^{+}]\ket{\epsilon_{k-1}^k}&=(E_0+1-\epsilon_0-k)\ket{\epsilon_{k-1}^k},   
\end{align}\end{subequations}
i.e., due to its action on the states $\ket{\epsilon_0^k}$ and $\ket{\epsilon_{k-1}^k}$ it is seen that $\conm{\ell_k^-}{\ell_k^+} |_{\mathcal{H}_{\text{new}}} \neq \mathbb{1}|_{\mathcal{H}_{\text{new}}}$. Despite this fact, we are going to apply the same displacement operator of Equation~\eqref{lineD} onto the ground state $\ket{\epsilon_0^k}\in\mathcal{H}_{\text{new}}$ to obtain
\begin{equation}
\ket{z_\text{new}^k}=C_z' D(z)\ket{\epsilon_0^k}=C_z' \sqrt{\Gamma(E_0-\epsilon_0)}\exp\left(-\frac{|z|^2}{2}\right)\sum_{j=0}^{k-1}\frac{(iz)^j}{j!}\frac{1}{\sqrt{\Gamma(E_0-\epsilon_0-j)}}\ket{\epsilon_j^k},  \label{csketk2}
\end{equation}
where $C_z'$ is a normalization constant chosen now to absorb the factor $\sqrt{\Gamma(E_0-\epsilon_0)}\exp(-|z|^2/2)$ so that
\begin{equation}
\ket{z_\text{new}^k}=C_z \sum_{j=0}^{k-1}\frac{(iz)^j}{j!}\sqrt{\frac{1}{\Gamma(E_0-\epsilon_0-j)}}\ket{\epsilon_j^k}. \label{csketk}
\end{equation}
Without lost of generality, the normalization constant $C_z$ can be chosen real and positive. Then
\begin{equation}
C_z=\left[ \sum_{j=0}^{k-1}\frac{|z|^{2j}}{(j!)^2}\frac{1}{\Gamma(E_0-\epsilon_0-j)}   \right]^{-1/2}.  
\end{equation}

These CS satisfy the following properties:
\begin{itemize}
\item \emph{Continuity of the labels.}
In order to check this property, we can see once again that
\begin{equation}
\norm{\ket{z'^k_\text{new}}-\ket{z^k_\text{new}}}^2=\inner{z'^k_\text{new}-z^k_\text{new}}{z'^k_\text{new}-z^k_\text{new}}=2\left[1-\text{Re}(\inner{z'^k_\text{new}}{z^k_\text{new}}) \right].  
\end{equation}
Let us define the complex function $C(a,b)$ as
\begin{equation}
C(a,b)= \left[\sum_{j=0}^{k-1} \frac{(a^*b)^j}{(j!)^2} \frac{1}{\Gamma(E_0-\epsilon_0-j)}\right]^{-1/2}.  
\end{equation}
Therefore
\begin{equation}
\inner{z'^k_\text{new}}{z^k_\text{new}}=\frac{C(z',z')C(z,z)}{C^2(z',z)},  
\end{equation}
which implies that in the limit $z' \rightarrow z$ it turns out that $\ket{z'^k_\text{new}} \rightarrow \ket{z^k_\text{new}}$. 

\item \emph{Resolution of the identity.}
Recall that this property requires the following expression to be satisfied
\begin{equation}
\int_\mathbb{C} \ket{z^k_\text{new}} \bra{z^k_\text{new}}\mu_3(z)\text{d}z=\mathbb{1}|_{\mathcal{H}_\text{new}},
\end{equation}
where $\mu_3(z)$ is a positive definite function to be found. If we substitute the expression $\ket{z^k_\text{new}}$ given by Equation \eqref{csketk}, express $z$ in polar coordinates, suppose that $\mu_3(z)=\mu_3(r)$, and integrate the angular variable we obtain
\begin{equation}
\mathbb{1}|_{\mathcal{H}_\text{new}}= 2\pi\sum_{j=0}^{k-1}\frac{\ket{\epsilon_j^k} \bra{\epsilon_j^k}}{(j!)^2\Gamma(E_0-\epsilon_0-j)}\int_0^\infty C_z^2 r^{2j+1}\mu_3(r)\text{d}r.  
\end{equation}
In order to simplify this equation, we introduce the function $f_3(r)$ as
\begin{equation}
\mu_3(r)=\frac{f_3(r)}{\pi C_z^2}, \label{csmu}
\end{equation}
in such a way that the following equation must be fulfilled
\begin{equation}
2\int_0^\infty r^{2j+1}f_3(r)\text{d}r=\Gamma^2(j+1)\Gamma(E_0-\epsilon_0-j).  
\end{equation}
With the change of variable $r^2=x$ and of index $j=s-1$ we obtain
\begin{equation}
\int_0^\infty x^{s-1}f_3(x)\text{d}x  =\Gamma^2(s)\Gamma(E_0+1-\epsilon_0-s)\equiv \mathcal{M}[f(x);s].  
\end{equation}
Now we need to find the {\it inverse Mellin transform}
\begin{equation}
f_3(x)=\mathcal{M}^{-1}[\Gamma^2(s)\Gamma(E_0+1-\epsilon_0-s);x]. \label{cseqf}
\end{equation}
It is possible to find several inverse Mellin transforms in tables, for example in Erd\'elyi's book \cite{Erd54}. In this case, the function $f_3(x)$ of Equation~\eqref{cseqf} turns out to be a Meijer $G$ function with $m=2,\, n=1,\, p=1,\, q=2,\, a_1=\epsilon_0-E_0,\, b_1=b_2=0$, i.e.,
\begin{equation}
f_3(r)=\MeijerG{2}{1}{1}{2}{\epsilon_0-E_0}{0,\, 0}{r^2}.  
\end{equation}
Moreover, in Erd\'elyi's book of transcendental functions \cite{Erd53} one can find some expressions for the Meijer $G$ function in terms of other special functions, in particular of the Whittaker function $W_{\kappa,\mu}(z)$, which in turn can be written in terms of the logarithmic solution of the confluent hypergeometric equation $U(a,c;z)$ \cite{AS72}. Then we have
\begin{equation}
f_3(r)=\Gamma^2(E_0+1-\epsilon_0) U(E_0+1-\epsilon_0,1;r^2).
\end{equation}
 
In order to prove the positiveness of $\mu_3(z)$ we will use again the generalized Parseval formula. In this way, if we choose $h^*(s)=\Gamma(s)\Rightarrow h(x)=\exp(-x)$, $g^*(s)=\Gamma(s)\Gamma(E_0+1-\epsilon_0-s)$ and use the following equation \cite{Erd54} 
\begin{equation}
\mathcal{M}^{-1}[\Gamma(\alpha+s)\Gamma(\beta-s)]=\Gamma(\alpha+\beta)x^\alpha(1+x)^{\alpha-\beta},  
\end{equation}
it is obtained
\begin{equation}
g(x)=\Gamma(E_0+1-\epsilon_0)(1+x)^{\epsilon_0-E_0-1}.  
\end{equation}
Using now the generalized Parseval formula from Equation \eqref{parseval2} it turns out that
\begin{align}
f_3(x)&=\int_0^{\infty}\Gamma(E_0+1-\epsilon_0)(1+xt^{-1})^{\epsilon_0-E_0-1}\exp(-t)t^{-1}\text{d}t  \nonumber  \\
&=\Gamma(E_0+1-\epsilon_0)\int_0^\infty t^{E_0-\epsilon_0}(t+x)^{\epsilon_0-E_0-1}\exp(-t)\text{d}t.  
\end{align}
If we replace this last result in Equation~\eqref{csmu} we obtain
\begin{equation}
\mu_3(r)=\frac{\Gamma(E_0+1-\epsilon_0)}{\pi C_z^2} \int_0^\infty t^{E_0-\epsilon_0}(t+r^2)^{\epsilon_0-E_0-1}\exp(-t)\text{d}t.
\end{equation}
Besides, taking into account that $E_0>\epsilon_0$, and that the domain of $r$ and $t$ is $[0,\infty)$ we can conclude that we have found, at least, one positive definite measure, i.e, the CS of Equation \eqref{csketk} do resolve the restriction of the identity operator in the subspace $\mathcal{H}_{\text{new}}$. 

\item \emph{Temporal stability.}
If we apply the evolution operator to a CS in the subspace $\mathcal{H}_{\text{new}}$ we obtain
\begin{align}
U(t)\ket{z^k_{\text{new}}}&=C_z \sum_{j=0}^{k-1}\frac{(iz)^j}{j!}\sqrt{\frac{1}{\Gamma(E_0-\epsilon_0-j)}}\exp(-iH_kt)\ket{\epsilon_j^k} \\
&=\exp(-i\epsilon_0t)C_z \sum_{j=0}^{k-1}\frac{[iz\exp(-it)]^j}{j!}\sqrt{\frac{1}{\Gamma(E_0-\epsilon_0-j)}}\ket{\epsilon_j^k} \\
&=\exp(-i\epsilon_0t)\ket{z\exp(-it)^k_{\text{new}}}\equiv \exp(-i\epsilon_0t)\ket{z^k_{\text{new}}(t)}.
\end{align}
We can see that, up to a global phase factor, a CS $\ket{z^k_{\text{new}}}\in \mathcal{H}_{\text{new}}$ always evolves into another CS in the same subspace.

\item \emph{Mean energy value.}
In order to find the mean energy value we use the explicit expression of the CS of Equation \eqref{csketk}, leading to:
\begin{equation}
\esp{H_k}_{z_\text{new}}=\bra{z_\text{new}^k}H_k\ket{z_\text{new}^k}
=\epsilon_0 +  C_z^2 \sum_{j=0}^{k-1}\frac{j |z|^{2j}}{(j!)^2}\frac{1}{\Gamma(E_0-\epsilon_0-j)}.
\end{equation}

\item \emph{State probability.}
We can easily calculate now the probability $p_j(z)$ that, for the system being in a CS $\ket{z^k_{\text{new}}}$, an energy measurement will give as a result the eigenvalue $\epsilon_j$, namely,
\begin{equation}
p_j(z)=|\inner{\epsilon_j^k}{z^k_\text{new}}|^2 =\frac{|z|^{2j}}{(j!)^2} \frac{C_z ^2}{\Gamma(E_0-\epsilon_0-j)} .  
\end{equation}

An example of the modulus squared of the wavefunction associated to a CS $\ket{z^k_{\text{new}}}$ is shown in Figure 4.
\end{itemize}

\begin{figure}[ht]
\centering \includegraphics[width=12cm]{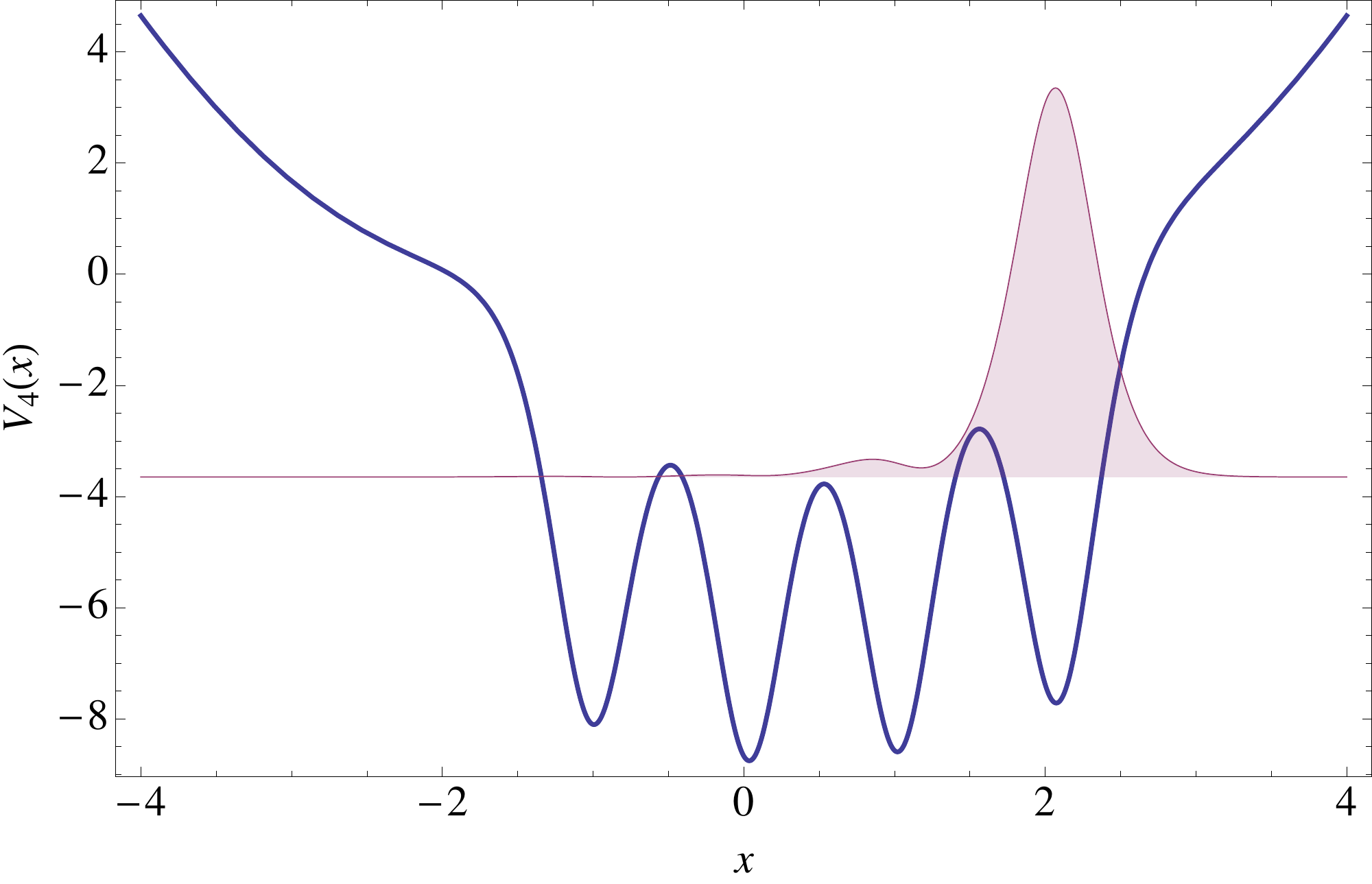}
\caption{The modulus squared of the wavefunction associated to a coherent state $\ket{z^k_{\text{new}}}$ in $\mathcal{H}_{\text{new}}$ with $z=1.5 e ^{-4.93 i}$ and mean energy value $-3.64945$ for a Painlev\'e IV Hamiltonian system obtained with $k=4$, $\epsilon_3=-2.8$, and $\nu_3=-0.9$.}
\label{Figure LinearizedCSnew}
\end{figure}

\section{Concluding remarks}
In this work we have studied the CS for a special kind of Hamiltonian systems which are connected with the Painlev\'e IV equation through a second-order polynomial Heisenberg algebra. First, we established the relation that the Painlev\'e IV equation hold with certain Hamiltonian systems. Then, using supersymmetric quantum mechanics the Painlev\'e IV Hamiltonian systems of our interest were constructed. Furthermore, the third-order ladder operators characteristic for these systems were employed to generate several families of coherent states. At the beginning we built the CS as eigenstates of the third-order annihilation operator, then as arising from acting the displacement operator involving these third-order ladder operators onto an extremal state. Finally, we used some linearized ladder operators for applying the corresponding displacement operator onto the extremal states in order to find new sets of CS. 

We must remember that the Painlev\'e IV Hamiltonian systems have two independent energy ladders, one semi-infinite starting from $E_0=1/2$, and one finite with $k$ levels which starts from $\epsilon_0$, where $k$ is the order of the SUSY transformation used to generate the potential. Thus, it is quite natural that the system is described by two orthogonal subspaces: one generated by the eigenstates associated to the energy levels of the harmonic oscillator, denoted as $\mathcal{H}_{\text{iso}}$, and another one generated by the eigenstates associated with the new levels, denoted as $\mathcal{H}_{\text{new}}$.

For the PIVCS which are eigenstates of the annihilation operator $l_k^-$, we were able to obtain a suitable set of CS only in the subspace $\mathcal{H}_{\text{iso}}$. For the PIVCS arising from the displacement operator which involves $l_k^\pm$, we have found a suitable set only in the complementary subspace $\mathcal{H}_{\text{new}}$. Finally, for the linearized PIVCS arising from the displacement operator which involves the linearized ladder operators $\ell_k^\pm$ we have found good sets of CS in both subspaces $\mathcal{H}_{\text{iso}}$ and $\mathcal{H}_{\text{new}}$. 

We must remark that the sets of CS which were found for the separated subspaces with different definitions are also good ones. Finally, 
some physical and mathematical properties of these families of coherent states were also studied. 

\section*{Acknowledgments}
The authors acknowledge the support of Conacyt, Project 152574. DB and ACA also acknowledge Conacyt fellowships 207672 and 207577.

\end{document}